\documentclass[11pt]{article}
\usepackage{graphicx}

\newcommand{\BABARPubYear}    {05}

\newcommand{\BABARConfNumber} {05/006}
\newcommand{\SLACPubNumber} {11329}
\newcommand{\LANLNumber} {0000}

\input pubboard/babarsym
\newcommand{\etal}      {\mbox{\textsl{et al.}}\xspace}
\newcommand{\forex}     {\mbox{\textsl{e.g.}}\xspace}
\newcommand{\ie}        {\mbox{\textsl{i.e.}}\xspace}

 \def\kg         {\ensuremath{K^{*} \gamma}\xspace}
 \def\xs         {\ensuremath{X_{s} }\xspace}
 
 \def\pizeta     {\ensuremath{\piz(\eta)}\xspace}
 \def\et     {\ensuremath{\eta}\xspace}

 \def\eg         {\ensuremath{E_{\gamma} }\xspace}

 \def\egcms      {\ensuremath{E^{*}_{\gamma}}\xspace}

 \def\mxs        {\ensuremath{m_{X_{s}} }\xspace}

 \def\mup        {\ensuremath{\mu_{\pi}}\xspace}
 \def\mb         {\ensuremath{m_{b} }\xspace}  
 
 \def\acp        {\ensuremath{A_{CP}}\xspace}

 \def\sig        {\ensuremath{\mathrm{S}}\xspace}
 \def\bkb        {\ensuremath{\mathrm{B}}\xspace}
 \def\bkc        {\ensuremath{\mathrm{C}}\xspace}

 \def\bsg        {\ensuremath{b \to s \gamma}\xspace}
 \def\bdg        {\ensuremath{b \to d \gamma}\xspace}

 \def\bxsg       {\ensuremath{B \to X_{s} \gamma}\xspace}
 \def\bxdg       {\ensuremath{B \to X_{d} \gamma}\xspace}

 \def\bkg        {\ensuremath{B \to \Kstar \gamma}\xspace}

\def\efmbrest   {\ensuremath{\langle \eg \rangle}\xspace}

\def\efmbreccms   {\ensuremath{\langle \egcms \rangle}\xspace}

\def\esmbrest   {\ensuremath{\langle \eg^{2} \rangle}\xspace}

\def\esmbreccms   {\ensuremath{\langle E_{\gamma}^{*2} \rangle}\xspace}

\def\varbrest   {\ensuremath{\esmbrest - {\efmbrest}^{2}}\xspace}
\def\varreccms   {\ensuremath{\esmbreccms - {\efmbreccms}^{2}}\xspace}
\def\aveDelta#1 {\ensuremath{\langle \Delta_{total}#1 \rangle}\xspace}

\def\cm         {\ensuremath{\mathrm{c.m.}}\xspace}

\long\def\inst#1{\par\nobreak\kern 4pt\nobreak
    {\it #1}\par\vskip 10pt plus 3pt minus 3pt}

\begin{document}
{\pagestyle{empty}

\begin{flushright}
\babar-CONF-\BABARPubYear/\BABARConfNumber \\
SLAC-PUB-\SLACPubNumber \\
hep-ex/\LANLNumber \\
\vspace{0.1in}
June 2005 \\
\end{flushright}
\par\vskip 5cm

\begin{center}
\Large \bf Results from the \babar\ Fully Inclusive Measurement of $B\to X_s\gamma$
\end{center}
\bigskip

\begin{center}
\large The \babar\ Collaboration\\
\mbox{ }\\
\today
\end{center}
\bigskip \bigskip

\begin{center}
\large \bf Abstract
\end{center}
We present preliminary results from a lepton-tagged fully-inclusive
measurement of \bxsg decays, where \xs is any strange hadronic state.
Results are based on a \babar\ data set of
88.5 million \BB pairs at the \Y4S resonance.  We present a
reconstructed photon energy spectrum in the \Y4S frame, and
partial branching fractions above minimum reconstructed photon energies of
1.9, 2.0, 2.1 and 2.2 GeV. We then convert these to measurements of
partial branching fractions and truncated first and second moments of the 
true photon energy distribution in the \B rest frame, above the same 
minimum photon energy values.
The full correlation matrices between the first and second moments are 
included to allow fitting to any parameterized theoretical calculation. 
We also measure the direct \CP asymmetry $\acp(B \to X_{s+d}\gamma)$ 
(based on the charge of the tagging lepton) above a reconstructed photon 
energy of 2.2 GeV.  \\

\vfill
\begin{center}
Contributed to the 
XXII$^{\rm st}$ International Symposium on Lepton and Photon Interactions at High~Energies, 6/30 --- 7/5/2005, Uppsala, Sweden
\end{center}

\vspace{1.0cm}
\begin{center}
{\em Stanford Linear Accelerator Center, Stanford University, 
Stanford, CA 94309} \\ \vspace{0.1cm}\hrule\vspace{0.1cm}
Work supported in part by Department of Energy contract DE-AC03-76SF00515.
\end{center}

\newpage
} 

\begin{center}
\small

The \babar\ Collaboration,
\bigskip

B.~Aubert,
R.~Barate,
D.~Boutigny,
F.~Couderc,
Y.~Karyotakis,
J.~P.~Lees,
V.~Poireau,
V.~Tisserand,
A.~Zghiche
\inst{Laboratoire de Physique des Particules, F-74941 Annecy-le-Vieux, France }
E.~Grauges
\inst{IFAE, Universitat Autonoma de Barcelona, E-08193 Bellaterra, Barcelona, Spain }
% ...

\end{center}\newpage

\section{Introduction} \label{sec:intro}

In the Standard Model the radiative decay \bsg\ proceeds via a loop
diagram, and is sensitive to possible new physics, with new heavy
particles participating in the loop
\cite{ref:new-physics}. Next-to-leading-order calculations for the
branching fraction give $\BR(\bxsg) = (3.57 \pm 0.30) \times
10^{-4}$\,($\eg > 1.6\gev$) \cite{ref:br-theory}, where \eg is the
photon energy in the \B rest frame; calculations to higher order are
currently underway~\cite{ref:theory-future}. Measurement of the \eg
spectrum from \bxsg decays, and in particular of its moments above
various minimum energies, allows the determination of the
heavy quark effective field theory parameters 
\mb and $\mup^{2}$, related to the
$b$-quark mass and momentum within the $B$ meson, respectively
\cite{ref:kn,ref:BBU,ref:neubert,ref:N}.  

This note reports on a fully inclusive analysis of data collected from
$\Y4S\to\BB$, where the photon from the decay of one \B meson is
detected, but the $X_s$ system is not reconstructed.  
Previous fully-inclusive measurements of $\bxsg$ have been
presented by the CLEO~\cite{ref:cleobsg} and BELLE~\cite{ref:bellebsg}
collaborations.  The alternative approach of summing a number of exclusive 
decays~\cite{ref:semi-inclusive} is more dependent on the \xs fragmentation 
model, and in particular on assumptions made as to the fraction of 
unmeasured final states.  In the fully-inclusive measurement we largely 
avoid such uncertainties, but at the cost of higher backgrounds, which need 
to be strongly suppressed.  Much of the non-\BB (continuum)
background is removed by requiring a high-momentum lepton tag; this
selects \BB events in which the non-signal \B~meson decays
semileptonically.  Continuum background is further suppressed by topological 
cuts.  Remaining continuum background is subtracted using off-resonance data 
recorded at an \epem center-of-mass (c.m.) energy just below that of the \Y4S.
Backgrounds from \BB events are estimated from a Monte Carlo 
simulation; for the most important backgrounds the simulation
is checked and corrected using data control samples.

We present preliminary results of this analysis, in particular the partial
branching fraction and the first and centralized second moments of the photon
energy distribution (\efmbrest and \varbrest, respectively) for each of
four minimum-\eg cuts in the \B rest frame: 1.9, 2.0, 2.1 and 2.2\gev. 
Hereafter we refer to the centralized second moment  as the ``second moment''.
These are obtained from the corresponding quantities for reconstructed 
photon energy \egcms in the \epem \cm frame (an asterisk denotes a
a \cm quantity) by applying corrections for resolution smearing and
the effect of imposing an experimental cut on \egcms rather than
\eg. The full correlation matrix between the first and second
moments is given to facilitate fitting to any parameterized
theoretical calculation.  The measurement does not distinguish \bxdg
events from
\bxsg events.  We assume the ratio of \bxdg to \bxsg is 
$|V_{td}/V_{ts}|^{2}$,  so that it constitutes $(4.0 \pm 1.6)\%$ of the
rate \cite{ref:cleobsg}, and correct for this at the end.  This assumption is
supported by newer theoretical calculations~\cite{ref:acptheory,ref:AAH}.
We also present a measurement of the \CP asymmetry in
the sum of \bxsg plus \bxdg, based on the charge of the tagging lepton. 
Pending a more detailed study of the energy spectrum, 
we do not yet perform the model-dependent extrapolation to the 
entire photon spectrum.

The results presented are based on data recorded with
the \babar\ detector~\cite{ref:babar-nim} at the PEP-II
asymmetric-energy \epem collider located at the Stanford Linear
Accelerator Center. The on-resonance data integrated luminosity is 
81.5\invfb, corresponding to 88.5 million \BB events. Additionally,
9.6\invfb of off-resonance data are used in the continuum background
subtraction. The \babar\ Monte Carlo simulation program, based on GEANT
4~\cite{ref:geant} and JETSET~\cite{ref:jetset}, is used to generate
samples of signal events, \BpBm and \BzBzb (excluding signal channels), \qqbar
($udsc$ continuum) and $\tau^{+}\tau^{-}$.
In order to minimize any possible experimenter's
bias, the spectrum of reconstructed photon energy \egcms from 1.9 to
2.9\gev in the on-resonance data was ``blinded'', \ie, not looked at until all
selection cuts were set and dominant \BB corrections determined.

\section{Signal Model and Backgrounds}\label{sec:model}

In order to compute signal (\bxsg) efficiency, we use the \mxs spectrum
computed by Kagan and Neubert (KN)~\cite{ref:kn}. 
This quantity has a 
one-to-one correspondence with the photon energy in the \B-meson 
rest frame via the kinematic relation
\begin{equation}
\label{eq:egamma}
   \eg = \frac{m_B^2 - m_{X_s}^2}{2m_B}\ .
\end{equation}
The KN spectrum is a smooth function of \mxs, \ie, it does not explicitly model
strange resonances in the mass spectrum. 
This assumption is not valid for the decay
$B\to K^*(892)\gamma$, so the spectrum needs to be modified to
describe it.  Higher-mass resonances are broad and
spaced closely enough, such that the smooth approximation of the KN
treatment is justified. In practice, a relativistic 
Breit-Wigner distribution modeling
the \bkg decay is used for $\mxs < 1.1\gevcc \, (\eg > 2.525 \gev)$ and the KN
spectrum is used above the cutoff. The \kg part is normalized
to the fraction of the KN spectrum below the \mxs cutoff.  The
KN model is parameterized in terms of two quantities, \mb (the
$b$-quark mass) and $\mup^2$ (a measure of the quark's Fermi momentum
inside the $B$ meson).  We chose a KN480 model, with
$\mb=4.80\gevcc$ and $\mup^2=0.3$ GeV$^2$, for studying systematic
effects on signal efficiency and for optimizing selection criteria.
For the latter purpose, an overall normalization was applied such that the 
\bkg component matches the previous \babar\ measurement~\cite{ref:ksg}.
In order to assess possible model-dependence of our results we use 
additional KN models with \mb between 4.55 and 4.80\gevcc, along with
a set of models based on the theory of ref.~\cite{ref:BBU}.

Backgrounds arise from continuum ($\epem\to\qqbar$, where
$q=udsc$, and \tautau) and \BB events. There are 
additional contributions from QED and hadronic two-photon processes, which 
are not simulated.
In \BB events, the main backgrounds
come from decays of \piz and $\eta$ mesons, with a smaller
contribution from hadrons (mostly antineutrons) that are
misidentified as photons. In continuum events, these 
backgrounds also contribute, along with initial state radiation (ISR),
where a high-energy photon is radiated from one of the colliding
\epm.  

\section{Event Selection}
\label{sec:evsel}

The \babar\ detector~\cite{ref:babar-nim} tracks charged particles with a 
five-layer double-sided silicon-strip detector and a forty-layer drift 
chamber placed inside a 1.5 T solenoidal magnet.  Electromagnetic 
showers are detected in a total-absorbtion calorimeter consisting of 
6580 CsI(Tl) scintillating crystals arranged hermetically in a projective
geometry.  Muons are indentified by resistive-plate chambers interleaved
in the flux-return yoke of the magnet, outside of the calorimeter

The distinctive feature of the signal decays is an isolated photon
with \cm energy from about 1.6\gev to 2.8\gev.
Our starting point is finding in the event at least one isolated photon 
candidate, \ie, a localized calorimeter energy deposit, with
$1.5<\egcms<3.5\gev$ (``photon acceptance'').
We then apply several categories of selection cuts, described in the
next few paragraphs:  hadronic event selection, photon quality criteria,
event shape cuts and tagging requirements.
Table~\ref{table:sel_eff} shows the efficiencies of the different categories
of cuts for signal and background event samples. 
In order to avoid regions where the expected signal would still be overwhelmed
by background (especially \BB background at low \egcms), our results 
will be quoted for several minimum-\egcms cuts of 1.9\gev or above.
The table illustrates the efficiencies for one particular choice.

Hadronic events are selected 
by requiring at least three reconstructed charged tracks and the 
second Fox-Wolfram moment $R_2^*<0.9$, calculated in the \cm frame using 
all detected particles. To reduce QED and two-photon backgrounds,
we use a cut on ``effective multiplicity,''
defined as the number of charged tracks plus half the number of 
photons with energy above 0.08\gev. We require $n_\mathrm{eff} \ge 4.5$,
a value that has essentially no impact on signal efficiency.

  Photon quality cuts are designed to suppress high-energy photon 
candidates which arise from backgrounds.  
We require that the polar angle of the photon satisfy
$-0.74 < \cos \theta_\gamma < 0.93$, to ensure that 
the photon shower is well contained in the calorimeter.
We suppress hadronic showers and \piz{}s with merged photon showers by 
requiring that the shower has a lateral profile consistent with a single photon.  
Photons that are consistent with originating from an identifiable \piz or
$\eta$ decay are explicitly vetoed. The optimal \pizeta veto requirements are 
energy-dependent, with a more open selection for $\egcms>2.3\gev$.
Finally, isolated photons are
selected by requiring that no energy (neutral or charged) be within 25
cm (approximately  seven times the  ${ \rm Moli\grave{e}re}$ radius for 
$CsI$)  of the photon, as measured at on the calorimeter's inside surface. 

Event shape variables are employed to exploit the difference in
topology of \BB events and continuum events. The former tend to be
spherical in shape, since the B mesons are produced nearly at rest in
the \Y4S rest frame, while continuum events tend to have a
more jet-like structure. We combine 19 different variables into a
single Fisher discriminant.  Eighteen of the quantities are
the sum of charged and neutral energy found in 10-degree cones (from 0
to 180 degrees) centered on the photon candidate direction. The photon
energy is not included in any of the cones. The nineteenth variable is 
the quantity $R_2^\prime/R_2^*$, where $R_2^\prime$ is
the second Fox-Wolfram moment calculated in the frame recoiling against the 
photon (which for ISR events is the \qqbar\ rest frame), using all
particles except the photon. The Fisher coefficients were
determined by maximizing the separation power between simulated signal
and continuum events. Cuts on 
the Fisher variable and $R_2^*$ $(< 0.55)$ remove 84\% of the 
remaining continuum background at this stage.

Lepton tagging is used to further reduce the backgrounds from
continuum events. About 20\% of \B mesons decay semileptonically to
either $e$ or $\mu$, and the large mass of the \B meson tends to impart
large momentum to its lepton daughter. Leptons from fragmentation in
continuum events tend to be at lower momentum. Since the tag lepton
comes from the recoil \B meson, it does not compromise the
inclusiveness of the \bxsg\ selection. 
Electrons are identified by requiring a well-measured track matched to
an energy deposit in the calorimeter consistent with an electromagnetic 
shower. Muons are identified by matching a well-measured track to
hits in the muon detector, with no more than a minimum-ionizing
energy deposit in the calorimeter.  The tag lepton is required to
have \cm momentum greater than 1.25\gevc for electrons and 1.5\gevc for
muons. Furthermore, making a cut on the angle between the photon
and the lepton, $\cos
\theta_{\gamma\ell} > -0.7$, removes more continuum background, in
which the lepton and photon candidates tend to be
back-to-back. Finally, we exploit the presence of a relatively
high-energy neutrino in semileptonic \B decays and place a cut on the
missing energy in the event in the \cm frame, $E^{*}_\mathrm{miss}>0.8\gev$. 

Virtually all of the tagging leptons arise from the decay $B\to
X_c\ell\nu$. We correct the rate of such events in our  simulation
based on the studies presented in
Ref.~\cite{ref:semileptonic}. The overall rate of $B\to
X_c\ell\nu$ is comprised of four decay modes, where the hadronic
system is  $D$, $D^*$, $D^{**}$ (higher mass resonances), and
non-resonant $D^{(*)}\pi$, respectively.  A scale factor was determined for each mode and
we have weighted our simulated  semileptonic decays accordingly. 
We also reweight  
the simulated form factors of the $D^*\ell\nu$ decay to those measured 
by \babar \cite{ref:DstFF}.

\begin{table}[htbp]
\hspace*{1.0cm}
\begin{center}
\renewcommand{\arraystretch}{1.2}
\begin{tabular}{|l|ccc|} \hline
            & $B\to X_s \gamma$ & \BB        & $udsc$ \\ \hline
Photon acceptance & 0.782       & 0.023      & 0.124  \\ \hline\hline   
Hadronic          & 0.937       & 0.979      & 0.826  \\
Photon quality    & 0.753       & 0.175      & 0.196  \\
$2.0 < \egcms < 2.7~\gev$
                  & 0.910       & 0.114      & 0.326  \\
Event shape       & 0.601       & 0.401      & 0.167  \\
Tagging           & 0.052       & 0.041      & $7.4\times 10^{-4}$ \\ \hline
Total selection & $2.0\times 10^{-2}$ & $3.2\times 10^{-4}$ & $6.5\times 10^{-6}$ \\ \hline
\end{tabular}
\end{center}
\vspace{-0.1in}
\caption{Selection efficiencies for signal and background events,
  broken down into different cut categories, as described in the
  text. The total selection efficiency is relative to the photon acceptance. 
  The signal model used is KN480 including $K^*\gamma$, as defined in
  the text. The error on the  total  signal efficiency from Monte Carlo statistics is $1.1\%$. }
\label{table:sel_eff}
\end{table}

The values for many of the cuts described above, along with the Fisher
coefficients, were optimized with all other cuts applied, using an
iterative technique.  In each case, the quantity maximized was
$\sig^2/(\sig + \bkb + \alpha\bkc)$, where \sig is the expected signal
from the KN480 model described above, \bkb is the simulated \BB background,
and \bkc is the simulated continuum background, each being the 
number of events expected for the
on-resonance data luminosity, and $\alpha$ is the ratio of 
total to off-resonance luminosity. The factor $\alpha$ takes
into account the fact that we later use off-resonance data 
to subtract the residual continuum background.

\begin{figure}[htb]
\begin{center}   
  \includegraphics[width=.6\textwidth]{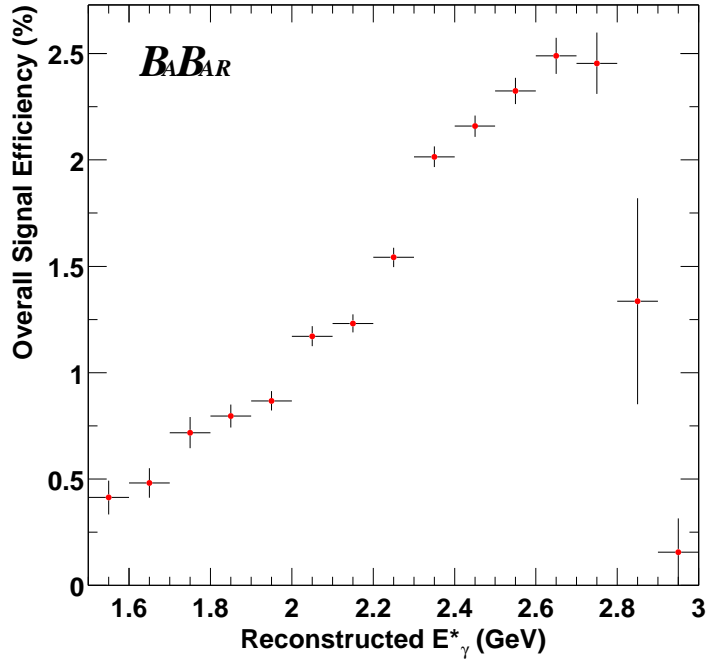}   
\end{center}
\vspace{-0.2in}
\caption{The overall signal efficiency in bins of \egcms (computed using
the KN480 model as defined in the text).}
\label{fig:efficiency} 
\end{figure}

Figure~\ref{fig:efficiency} shows the overall signal efficiency in bins 
of \egcms. The \egcms dependence is a result of the optimization 
procedure, in particular the training of the Fisher discriminant. The sum of 
the 18 energy-flow variables combined in the Fisher discriminant is correlated 
with the energy of the photon in the \B rest frame. Without the Fisher 
selection the continuum background 
decreases by a factor of two over the 2.0 to 2.7\gev range 
of \egcms used for optimization, while the expected 
signal increases rapidly betewen 2.0 and about 2.4\gev. 
Therefore, the optimization  
suppresses background more strongly with decreasing
 \egcms, leading to a corresponding reduction of the efficiency. The more 
stringent \pizeta vetoes applied in events with $\egcms<2.3\gev$ 
also contribute to the energy dependence. The variation of the efficiency 
with \egcms does not introduce a significant dependence on the multiplicity 
of the \xs system, which would lead to uncertainties
related to fragmentation modeling. However, the 
variation of the efficiency with \egcms does lead to a dependence on 
the modeling of the photon energy spectrum. The dominant effect is that 
reducing the assumed value of \mb in a particular model 
leads to reduction in the mean \egcms, which 
in turn yields a smaller selection efficiency. We 
improve the estimation of the signal efficiency and reduce the 
model-dependent uncertainty with a bootstrapping procedure 
described in section~\ref{sec:pbf}. 

\begin{figure}[htb]
\begin{center}   
  \includegraphics[width=.6\textwidth]{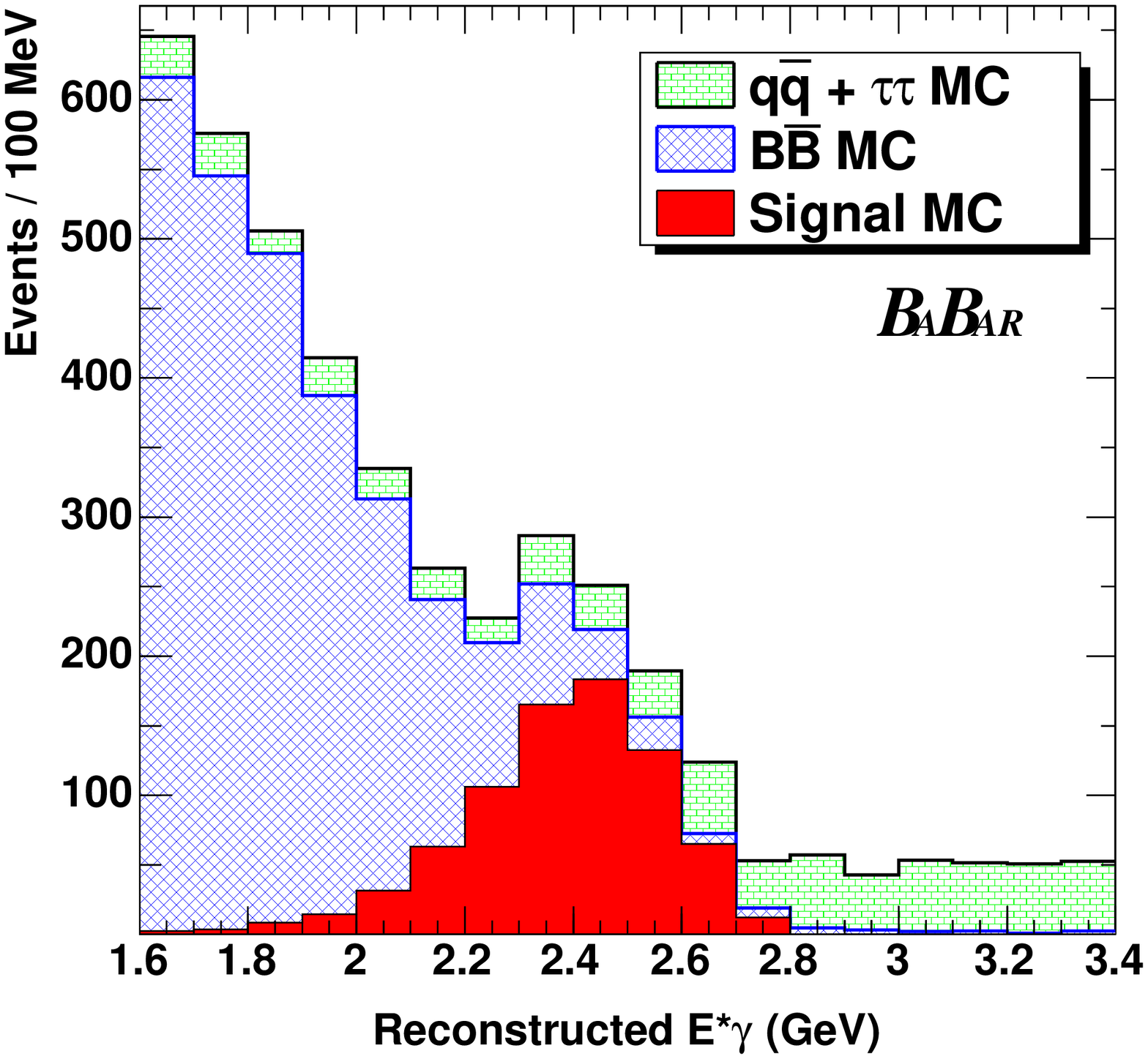}   
\end{center}
\vspace{-0.2in}
\caption{Monte Carlo simulation of
the \egcms distribution for \bxsg, continuum
(\uubar, \ddbar, \ssbar, \ccbar, \tautau) and \BB background after all 
selection criteria (except that on \egcms itself) have been applied.}
\label{fig:mcexpect} 
\end{figure}

 The simulated \egcms distribution of signal, continuum and \BB background 
events after all selection criteria is shown in Fig.~\ref{fig:mcexpect}. Note that the
continuum Monte-Carlo does not include QED backgrounds present in the data.
The actual  continuum background is subtracted using  the off-resonance data 
and is the dominant source of  statistical uncertainty on the measurements.

\section{\BB Backgrounds}
\label{sec:bb}
As can be seen in Figure~\ref{fig:mcexpect}, 
the \BB background dominates for
$\egcms < 2.4\gev$. We present in Table~\ref{tab:bb} the various
sources of background coming from \BB events, as predicted by our
Monte Carlo simulation. 
\begin{table}[htbp]	
 \begin{center}
 \begin{tabular}{|l|l|c|} \hline 
                          &           & Fraction of \BB background \\
  Truth-match             & Parentage & ($2.0 < \egcms < 2.7\gev$)  \\ 
  \hline			     
  Photon             	  & \piz      & 0.640 \\
  			  & $\eta$    & 0.174 \\
  			  & $\omega$  & 0.024 \\
  			  & $\eta'$   & 0.011 \\
  			  & FSR       & 0.007 \\
  			  & \jpsi     & 0.008 \\
  			  & Other     & 0.001 \\ 
  			  & Total     & 0.865 \\ \hline
  \piz (merged)       	  & Any       & 0.001 \\
  Electron           	  & Any       & 0.036 \\
  \antineutron (\neutron) & Any       & 0.077 \\
  \antiproton (\proton)   & Any       & 0.005 \\
  \KL                     & Any       & 0.001 \\
  \pipm or \Kpm           & Any       & 0.001 \\ \hline
  Unidentified            &           & 0.015 \\ \hline
 \end{tabular}
\end{center}
\vspace{-0.2in}
\caption{Breakdown of Monte Carlo simulated  \BB backgrounds by high-energy photon origin for $2.0 < \egcms < 2.7\gev$. 
  These fractions are for the combined
  \BzBzb and \BpBm Monte Carlo samples before tagging cuts.}
 \label{tab:bb}
\end{table}

About 80\% of the background comes from photons from \pizeta\ decays,
for both the nominal signal region ($2.0 < \egcms < 2.7\gev $) and a
control region ($1.6 < \egcms < 1.9\gev $). Other significant sources
are hadrons in the signal region ($\sim 8$\%) and electrons in the
control region ($\sim 9$\%). We have performed dedicated studies 
of data control samples for these sources, 
to verify and/or correct the Monte Carlo predictions. 

The level of the largest \BB background component, \pizeta\ decays,
has been measured directly from the data. We explicitly measure the
yield of \piz and \et decays in bins of $E^*_{\piz}$ or
$E^*_{\eta}$ in on-resonance data, off-resonance data and \BB
simulation samples, by fitting the \gaga mass distributions. This allows us to
derive correction factors versus \pizeta energy for the \pizeta rates 
predicted by the simulation. 
These corrections are then applied to our \BB simulation when estimating the \BB
background passing the $\bxsg$ selection. The \pizeta analysis uses the same
selection criteria as for the $\bxsg$ analysis, with three exceptions: 1)
the \pizeta vetoes are not applied, 2) the requirement on the photon
energy is reduced to $\egcms > 1.0\gev$ to increase 
statistics, and 3) the tagging momentum cuts are loosened to 1.0\gevc for
electrons and 1.1\gevc for muons, again to increase the available
statistics. We apply an additional small correction to
account for differences in the performance of the \pizeta\ veto
in data and \BB simulation. 

The background from hadrons faking photons in \BB events consists primarily 
of $\antineutron$\,s with a small contribution from $\antiproton$\,s. We have 
corrected the \BB simulation for both the $\antineutron(\antiproton)$ 
production and $\antineutron(\antiproton)$ response in the calorimeter.
First, an  inclusive measurement of \proton and
\antiproton\ production, 
is used to correct the simulation prediction for both \antiproton\ and \antineutron\
production. A conservative 50\% systematic uncertainty is assigned to account for
the fact that we are using a \antiproton\ measurement to correct
\antineutron\ production.  Additionally, $\bar \Lambda \to \antiproton
\pi^+$ decays are studied both in real data and simulation to compare the
response of the calorimeter to \antiproton{}\,s.  We
find a significantly smaller number passing the \egcms and\
lateral-shower-profile cuts 
in the data than that expected from the simulation. The
total correction factor, then, is a product of the production
correction and the efficiency correction, determined in bins of
\egcms. 

Electrons can cause a background to the photon sample if the charged
track is not reconstructed or matched to the electromagnetic shower.
(Bremsstrahlung photons are an additional source of 
background from $e^{\pm}$, which is well modeled in the simulation.)
The electron tracking efficiency was measured in data by selecting one
track of a Bhabha event and then measuring the probability of
reconstructing the other electron track.  A further
correction is made to account for the fact that the environment of the
control sample (Bhabha events) has significantly lower multiplicity on
average than the
actual data sample (hadronic events).

We have also derived corrections for the very small number of background 
photons from $\omega$ and $\etapr$ decays. 
The corrections to the modeling of $\B\to\X_c\ell\nu$ 
described in section~\ref{sec:evsel} are applied to the \BB simulation.
After all corrections, we check our estimate of the \BB\ background by
considering the low energy control region $1.6 < \egcms < 1.9\gev$,
where we expect few photons from \bxsg decays. In this region we
observe $1790 \pm 64$ events after continuum subtraction compared 
to an expection of $1667 \pm 54$ events, \ie  
\begin{equation} 
  \mathrm{Observed} - \mathrm{Expected} = 123 \pm 64 (\mathrm{stat.}) 
  \pm 54 (\mathrm{syst.}) 
\end{equation}
The significance of the deviation is $1.5\sigma$. Note that for this check
we have not subtracted any signal contribution, which for the KN480 and
KN465 models defined above would be approximately 17 and 39 events,
respectively. These would reduce the significance of the deviation to
$1.3\sigma$ and $1.0\sigma$, respectively. 

\section{Results}

\subsection{Partial Branching Fractions}
\label{sec:pbf}

\begin{table}[!b]
 \begin{center}
  \begin{tabular}{|l|c|} \hline
   \egcms (GeV) & Tagged Signal Yield (events) \\ \hline
   1.9 to 2.7 & $  1042 \:\pm\: 84 \:\pm\: 62 $  \\
   2.0 to 2.7 & $ \ 992 \:\pm\: 77 \:\pm\: 53 $  \\
   2.1 to 2.7 & $ \ 895 \:\pm\: 72 \:\pm\: 45 $  \\
   2.2 to 2.7 & $ \ 758 \:\pm\: 66 \:\pm\: 40 $  \\
   \hline
  \end{tabular}
  \caption{Extracted tagged signal yields in various ranges of \egcms,
   uncorrected for efficiency.  The first
    error is statistical, the second is from \BB systematics.}
  \label{tab:yields}
 \end{center}
 \vspace{-0.2in}
\end{table}

\begin{figure}[htb]
\begin{center}   
  \includegraphics[width=.6\textwidth]{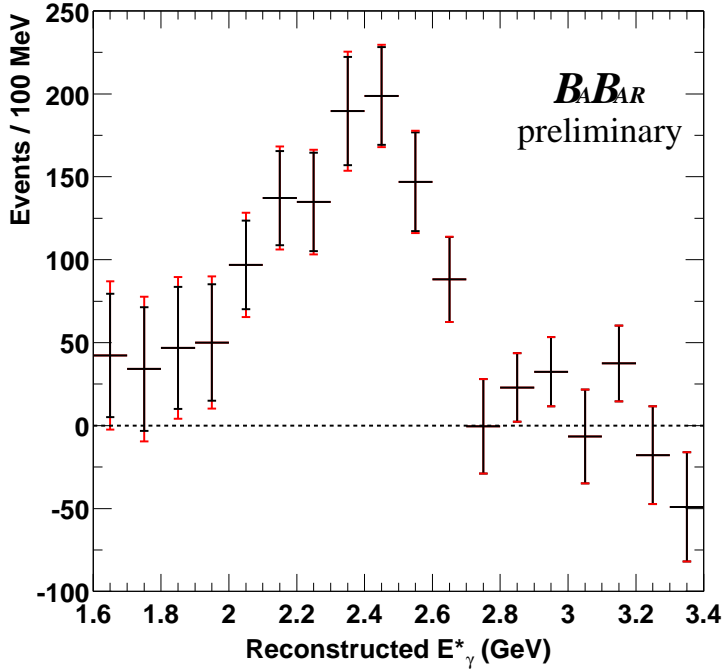}   
\end{center}
\vspace{-0.2in}
\caption{Photon energy spectrum after background subtraction, 
  uncorrected for efficiency.
  Error bars include statistical (dominant) and 
  \BB systematic uncertainties, added in quadrature.}
\label{fig:spectrum} 
\end{figure}

The extracted signal yield, \ie, the on-resonance yield minus the sum of
the scaled
off-resonance yield and the corrected \BB simulation prediction, is given in
Table~\ref{tab:yields} for each of four different energy ranges.  The 
corresponding raw photon energy spectrum, uncorrected for efficiency, is 
shown in Figure~\ref{fig:spectrum}.  The region $2.9 < \egcms < 3.4\gev$ 
serves as a control region for the continuum subtraction. In this
region we observe $390 \pm 20$ events compared to an expectation 
from off-resonance data of
$391 \pm 57$ events; a difference of $-1 \pm 61$ events.

The model-dependence of the
efficiency can be reduced by a bootstrapping method, whereby we
initially assume a model (\forex, KN465) and use it to calculate the
mean of the efficiency-corrected energy spectrum between 2.0 and 2.7\gev. 
This is compared to the corresponding mean predicted by several KN
models with different values of \mb.  An iterative procedure is used to
determine which model best matches the measured mean and to then assign
a model-dependent uncertainty to the efficiency corresponding to the
uncertainty on the measurement of the mean.  The
relation between efficiency and mean energy is approximately linear,
with a best match close to a KN460 model (with $\mup^2 = 0.603$).
We have carried out the
same studies for twelve different parameterizations of the Benson, Bigi and 
Uraltsev (BBU)~\cite{ref:BBU} calculation, using the same prescription to 
include a \Kstar component as for KN.  We varied \mb between
4.45 and 4.75\gev and $\mup^2$ between 0.25 and $0.65\gev^2$. (Note that these
two parameters are defined differently than their KN counterparts.)
We set $\mu_G^2$ (the matrix element of the chromomagnetic operator,
responsible for the hyperfine splitting between \B and $B^*$ mesons)
to either 0.35 or $0.27\gev^2$.  
All the BBU results for the signal efficiency versus mean energy lie on the same 
line as for the KN model, confirming the model-independence of the
bootstrapping procedure.  Thus the KN460 efficiency is used to determine the
partial branching fractions (PBFs) according to:
\begin{equation}
 \BR(\bxsg, \egcms \mathrm{\ in\ range}) = \frac{\mathrm{Events\ in\ range}}
 {2 \times N_{\BB} \times\eps_{\mathrm{KN460}}} \ .
 \label{eq:pbf}
\end{equation}

\begin{figure}[htb]
\begin{center}   
  \includegraphics[width=.6\textwidth]{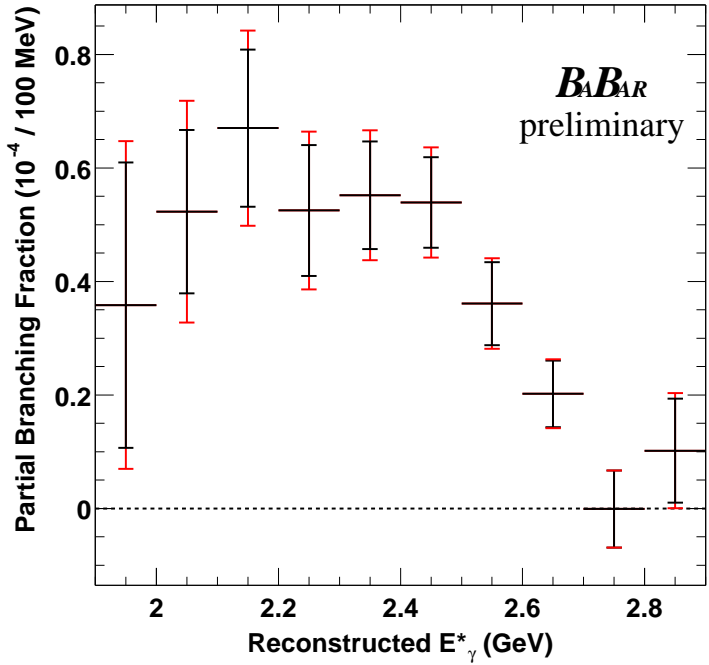}
\end{center}
\vspace{-0.2in}
\caption{Efficiency-corrected photon energy spectrum for the extracted 
signal, shown only for the originally-blinded range of reconstructed
energy (note the range 2.7-2.9\gev is not used to measure the branching fractions 
or moments).  
The small error bar is statistical only. The larger error bar also includes 
\BB and other systematic uncertainties and a model-dependence uncertainty,
all in quadrature.  There are significant correlations among 
non-statistical uncertainties for different bins.}
\label{fig:effcorrspec} 
\end{figure}

Figure~\ref{fig:effcorrspec}
shows the efficiency-corrected spectrum using the KN460 model.  
We have taken a conservative approach of assigning to each bin a
model-dependence uncertainty corresponding to the maximum efficiency
deviation, among KN455 to KN480 models and the 12 BBU models,
from the KN460 value.  Note that these models span a range of mean energy
up to $2\sigma$ away from our mesured value.

To make the results directly comparable to theoretical predictions, we
correct for the fact that the cut in photon energy is made
on reconstructed \egcms in the \Y4S (\cm) frame rather than on
true \eg in the \B rest frame.  The latter energy has been
smeared by a Doppler shift and by the (asymmetric) calorimeter resolution.
(Doppler smearing arises from the Lorentz transformation from the \B
rest frame to the \Y4S frame, given our ignorance of the direction of the \B.)
We compute the correction factor $\alpha_{cut}$ for the KN and BBU models and
find that it has minimal model-dependence.  
Table~\ref{table:partial_bfs_corrected} shows PBFs
with corrections applied, along with the
statistical, systematic and model-dependent errors.  For the corrected
PBFs the latter includes two correlated contributions:  the
model-dependent efficiency uncertainty noted above, already applied to
the measured PBFs, and the uncertainty on $\alpha_{cut}$.   

We studied many sources of systematic uncertainty, and here note
the more significant.
The uncertainty on the \BB background subtraction is shown in
Table~\ref{tab:yields}, and amounts to 5.5\% for 2.0 to 2.7\gev.
It comes mostly from the statistical uncertainties on the
correction factors derived from the \pizeta\ control sample.  Other
systematic effects total 6.4\% in quadrature.  Of this,
3.3\% is the uncertainty on photon selection,
dominated by a 2.5\% uncertainty on photon efficiency
(determined from $\piz$s in $\tau$ decays) 
and 2\% for the photon isolation cut.  It also includes
allowance for uncertainties in photon energy scale and
resolution, and in the photon lateral shape cut efficiency, derived mainly from
data from the \babar\ \bkg analysis and 
photons from virtual Compton scattering.  
The efficiency of the event shape cuts was studied using a \piz control
sample to compare distributions of the Fisher discriminant between
data and simulation, resulting in an uncertainty of 3.0\%.  A small
sensitivity to details of \xs fragmentation implies, for the adjustments 
determined in the \babar\ semi-inclusive analysis,
an additional uncertainty of only 1.4\%.  A 2.2\% uncertainty is
assigned for lepton identification, and
3.0\% for the uncertainties on the semileptonic corrections.

\begin{table*}[htbp]
\hspace*{1.0cm}
 \begin{center}
  \begin{tabular}{|c|c|c|c|c|} \hline
   E range      & Measured PBF ($\times 10^{-4}$)   &                  &                   &  Corrected PBF ($\times 10^{-4}$)   \\ 
   GeV          & for reco \egcms range, \cm frame  & $\alpha_{cut}$   & $\alpha_{d/s}$    &  for true \eg range, \B frame       \\ \hline
   1.9 to 2.7   & $3.64 \pm 0.29 \pm 0.33 \pm 0.22$ & $1.05 \pm 0.02$  & $0.96\pm 0.016$   &  $3.67 \pm 0.29 \pm 0.34 \pm 0.29$  \\
   2.0 to 2.7   & $3.32 \pm 0.26 \pm 0.28 \pm 0.16$ & $1.07 \pm 0.02$  & $0.96\pm 0.016$   &  $3.41 \pm 0.27 \pm 0.29 \pm 0.23$   \\
   2.1 to 2.7   & $2.84 \pm 0.23 \pm 0.23 \pm 0.11$ & $1.09 \pm 0.02$  & $0.96\pm 0.016$   &  $2.97 \pm 0.24 \pm 0.25 \pm 0.17$   \\
   2.2 to 2.7   & $2.23 \pm 0.19 \pm 0.18 \pm 0.06$ & $1.13 \pm 0.03$  & $0.96\pm 0.016$   &  $2.42 \pm 0.21 \pm 0.20 \pm 0.13$   \\ \hline
  \end{tabular}
 \end{center}
 \vspace{-0.1in}
\caption{Partial branching fractions (PBF{}s, preliminary) 
 corrected to the \B rest
 frame for four different reconstructed-\egcms ranges. The
 uncertainties shown for the PBF{}s are statistical, systematic and
 model-dependent. The ``Measured'' values are in terms of
 \egcms in the \cm frame.  The correction factor $\alpha_{cut}$
 converts the measured branching fraction for an $\egcms$ cut to a
 result in the \B rest frame with an equal $\eg$ cut, while the factor
 $\alpha_{d/s}$ removes the $(4.0 \pm 1.6)\%$ \bxdg component. For the
 last column, the error on $\alpha_{cut}$ is added linearly to the 
 fractional model-dependence error, while the error on $\alpha_{d/s}$ 
 is added in quadrature to the systematic error.}
 \label{table:partial_bfs_corrected}
\end{table*}

\subsection{Truncated Moments}
As noted above, the truncated moments of the photon energy
spectrum are of interest because they can be related to
parameters of heavy quark theory. 
The measured first moment $\efmbreccms$ and second moment $\varreccms$
are computed by averaging over the efficiency-corrected photon energy
spectrum in 100\mev bins.  We do this for four different energy
ranges, extending from 1.9, 2.0, 2.1 or 2.2\gev to 2.7\gev.
The bin-by-bin efficiencies derived from 
the KN460 model are used to correct the yields. 

The observed moments are derived from the measured \egcms spectrum 
but the moments computed in theoretical calculations involve the 
true \eg. Thus we need to correct
the former for Doppler and calorimeter-resolution smearing.  
For a spectrum without energy cuts, the total shift of the first 
(second) moment is the sum of the first (second) moments of the 
smearing functions, about $+0.005\gev \,(+0.007\gev^{2})$ for 
Doppler smearing plus $ -0.040\gev \,(+0.014\gev^{2}$) for 
the calorimeter.  The latter corrections are reduced when the spectrum 
is truncated by a minimum energy cut, \forex, to 
$-0.027\gev (-0.007\gev^{2})$ for a typical BBU model 
($\mb=4.6\gev,\mup^{2}=0.45\gev^{2}$)
with a cut of 2.0\gev.
An additional small shift accounts for cutting on
the reconstructed \egcms in the \Y4S frame instead of on the true
\eg in the \B rest frame.  
For a given signal model, Monte Carlo simulation provides
the total shift $\Delta_{total}$ needed to obtain the first moment \efmbrest 
from the measured  first moment \efmbreccms  for the same numerical cut value; likewise
for the second moment.

\begin{table*}[t]
\begin{center}
\begin{tabular}{|c|c|c|c|} \hline
  $E$ range  &   $\efmbreccms \pm(stat)\pm(sys)\pm(model)$ &    $ \Delta_{total}$    &  $\efmbrest \pm(stat)\pm(sys)\pm(model)$ \\
  GeV        &   \gev (with cut on \egcms)                 &         \gev            &  \gev (with cut on \eg)                  \\ \hline
  1.9 to 2.7 &   $2.270 \pm 0.025 \pm 0.017 \pm 0.005$     &    $ 0.018 \pm 0.007$   &  $2.288 \pm 0.025 \pm 0.017 \pm 0.012$   \\
  2.0 to 2.7 &   $2.304 \pm 0.016 \pm 0.010 \pm 0.005$     &    $ 0.012 \pm 0.007$   &  $2.316 \pm 0.016 \pm 0.010 \pm 0.012$   \\
  2.1 to 2.7 &   $2.350 \pm 0.014 \pm 0.007 \pm 0.003$     &    $ 0.005 \pm 0.007$   &  $2.355 \pm 0.014 \pm 0.007 \pm 0.010$   \\
  2.2 to 2.7 &   $2.412 \pm 0.012 \pm 0.005 \pm 0.002$     &    $-0.005 \pm 0.006$   &  $2.407 \pm 0.012 \pm 0.005 \pm 0.008$   \\ \hline
\end{tabular}
\end{center}
\vspace{-0.1in}
\caption{Measured first moments (preliminary) for various energy ranges, 
then corrected for 
resolution smearing and for cutting on the \B rest frame \eg.  The systematic
errors have two sources:  \BB background uncertainties, with their
full covariance matrix taken into account, and the uncertainty in the
photon energy scale.  The model-dependence error on $\efmbreccms$ is
correlated with that on $\Delta_{total}$, 
so these are combined linearly for the last 
column.  The \efmbrest values in that column can be compared to theoretical
predictions. Note that $\approx 4\%$ of \bxdg is included, but its first
moments are not expected to differ significantly from those for \bxsg.}
\label{table:firstmom_datacorrected}
\vspace{0.1in}
\end{table*}
 
\begin{table*}[t]
\begin{center}
\begin{tabular}{|c|c|c|c|} \hline
  $E$ range  &   $\varreccms \pm stat\pm sys$    &    $ \Delta_{total}$     &  $\varbrest \pm stat\pm sys\pm model$        \\
  GeV        &   $\gev^{2}$ (with cut on \egcms) &    $\gev^{2}$            &  $\gev^{2}$ (with cut on \eg)                \\ \hline
  1.9 to 2.7 &   $0.0403 \pm 0.0049 \pm 0.0022$  &    $ -0.0075 \pm 0.0025$ &  $0.0328 \pm 0.0049 \pm 0.0023 \pm 0.0025$   \\
  2.0 to 2.7 &   $0.0326 \pm 0.0026 \pm 0.0009$  &    $ -0.0060 \pm 0.0020$ &  $0.0266 \pm 0.0026 \pm 0.0010 \pm 0.0020$   \\
  2.1 to 2.7 &   $0.0246 \pm 0.0019 \pm 0.0005$  &    $ -0.0055 \pm 0.0015$ &  $0.0191 \pm 0.0019 \pm 0.0006 \pm 0.0015$   \\
  2.2 to 2.7 &   $0.0161 \pm 0.0014 \pm 0.0003$  &    $ -0.0045 \pm 0.0005$ &  $0.0116 \pm 0.0014 \pm 0.0004 \pm 0.0005$   \\ \hline
 \end{tabular}
\end{center}
\vspace{-0.1in}
\caption{Measured second moments (preliminary) for various energy ranges, 
then corrected for 
resolution smearing and for cutting on the \B rest frame \eg.  The
errors listed with $\Delta_{total}$ represent model-dependence.
The systematic errors have two sources:  \BB background uncertainties, 
with their full covariance matrix taken into account, and (after
the correction) the uncertainty 
in the modeling of the photon energy resolution.  The final corrected value 
for an \eg cut can be compared to
theoretical predictions.}
\label{table:secondmom_datacorrected}
\vspace{-0.1in}
\end{table*}

Tables~\ref{table:firstmom_datacorrected} 
and ~\ref{table:secondmom_datacorrected} show the observed and corrected 
first and second moment values for four different minimum photon energy
cuts.  Using 50\mev bins instead of 100\mev bins changes the 
first moment  by less than 0.003\gev (0.001\gev) and the second moment by less 
than $0.001\gev^{2} \, (0.0004\gev^{2})$ for the 1.9\gev cut (other cuts).
The dominant systematic uncertainty  arises from the \BB background
subtraction. In addition the uncertainty in the photon energy scale
and the modeling of photon energy resolution cause smaller systematic
errors in the first and second moments, respectively.  The
model-dependent uncertainty on the first moment  is based on redoing
the computation using the efficiency for each KN or BBU model that
matches the nominal first moment within its uncertainties. The second
moment is sensitive to the modeling of the low-energy tail of the
photon energy resolution but not to the efficiency model or to the
uncertainty in the photon energy scale.
    A systematic uncertainty for the photon energy resolution is
determined by comparing the asymmetric photon resolution function
measured using a sample of events from virtual Compton scattering (in
which the true photon energy can be determined by a kinematic
constraint) with the corresponding function for simulated events.  The
uncertainty varies from $0.0005 \gev^{2}$ for $\egcms > 1.9 \gev$ to
$0.0002 \gev^{2}$ for $\egcms > 2.2 \gev$ and is added in quadrature
to the systematic uncertainty arising from the \BB background
systematic.  We compute the
corrections applied to the moments for the full 
range of KN and BBU models considered, in order to better represent 
possible model-dependence.
The tabulated $\Delta_{total}$ is the average value of the maximum and
minimum corrections for these models, and its uncertainty 
is taken to be half the value of the difference
between the maximum and minimum. The model-dependent uncertainties on
\efmbreccms and $\Delta_{total}$ are correlated, and hence added linearly.

Finally Tables~\ref{table:moment_correlations_stat} through
~\ref{table:moment_correlations_model} show the full
correlation matrices between the eight measured moments 
for statistical, systematic, and model dependence errors, respectively. 
Statistical and systematic covariance matrices are computed using
standard error propagation, starting from the measured spectrum in energy
bins.  For systematics, this incorporates the bin-to-bin covariance matrix
that arises from the \BB subtraction, with the fully-correlated covariance
matrices arising from the energy-scale uncertainty (first moments only)
and calorimeter resolution (second moments only) added at the end.

The model correlations take into account the variation of both the
efficiency and the $\Delta_{total}$ corrections among different
models.  For each of the models (indexed $k$) and each of the eight
measured first and second moments $M_{j=1,8}$ we define
$\Delta_{res/cut}^{k}(M_{j})$ as the actual correction for that model.
We also define $\Delta_{eff}^{k}(M_{j})$ as the computed value of
the first or
second moment using a particular model minus its value using efficiencies
from the nominal KN460 model.  Letting
\[ \Delta_{total}^{k}(M_{j}) \equiv \Delta_{cut/res}^{k}(M_{j}) + \Delta_{eff}^{k}(M_{j}) \ , \]
we compute the covariance of $\Delta_{total}^k(M_{j})$ between each pair of moments
by averaging  over the  models, and thence derive the correlation matrix.
Note that this approach, which treats the predictions of the different
models as a statistical distribution, would yield smaller model-dependence
errors than the ones we quote in Tables~\ref{table:firstmom_datacorrected} 
and~\ref{table:secondmom_datacorrected}; but we
use this result solely for the correlation matrix, not for the errors
themselves.  

\begin{table*}[htbp]
 \begin{center}
  \begin{tabular}{|rc|cccccccc|}
   \hline
                                 &&  M  &  M  &  M  &  M  &  V  &  V  &  V  &  V  \\
   Quantity & Min. $\eg$  & 1.9 & 2.0 & 2.1 & 2.2 & 1.9 & 2.0 & 2.1 & 2.2 \\ \hline
   Mean     & $1.9\gev$ & 1.0000 & 0.5172 & 0.3548 & 0.2265 &-0.6110 & 0.0077 & 0.0821 & 0.1008  \\ 
   Mean     & $2.0\gev$ &        & 1.0000 & 0.6838 & 0.4326 & 0.2285 &-0.0008 & 0.1375 & 0.1645  \\
   Mean     & $2.1\gev$ &        &        & 1.0000 & 0.6260 & 0.4220 & 0.5660 & 0.1650 & 0.1884  \\     
   Mean     & $2.2\gev$ &        &        &        & 1.0000 & 0.4528 & 0.7568 & 0.7383 & 0.2113  \\     
   Variance & $1.9\gev$ &        &        &        &        & 1.0000 & 0.4626 & 0.3486 & 0.2520  \\  
   Variance & $2.0\gev$ &        &        &        &        &        & 1.0000 & 0.6966 & 0.4887  \\
   Variance & $2.1\gev$ &        &        &        &        &        &        & 1.0000 & 0.6709  \\
   Variance & $2.2\gev$ &        &        &        &        &        &        &        & 1.0000  \\
   \hline
  \end{tabular}
   \caption{Correlation matrix (preliminary) for statistical errors on the 
   truncated first and
   second moments (four each, for different minimum $\eg$ cuts).  This
   matrix is symmetric, so redundant entries are not shown. Column labels
   are shorthand for the identical row labels.}
 \label{table:moment_correlations_stat}
 \vspace{-0.2in}
 \end{center}
\end{table*}

\begin{table*}[htbp]
 \begin{center}
  \begin{tabular}{|rr|cccccccc|}
   \hline
                                 &&  M  &  M  &  M  &  M  &  V  &  V  &  V  &  V  \\
   Quantity & Min. \eg  & 1.9 & 2.0 & 2.1 & 2.2 & 1.9 & 2.0 & 2.1 & 2.2 \\ \hline
   Mean     & $1.9\gev$ & 1.0000 & 0.7841 & 0.6076 & 0.4622 &-0.7552 &-0.3862 & -0.1723 & -0.0942  \\
   Mean     & $2.0\gev$ &        & 1.0000 & 0.8007 & 0.6311 &-0.2515 &-0.4753 & -0.2298 & -0.1451  \\   
   Mean     & $2.1\gev$ &        &        & 1.0000 & 0.7886 &-0.0320 & 0.0202 & -0.2903 & -0.2115  \\
   Mean     & $2.2\gev$ &        &        &        & 1.0000 & 0.0489 & 0.1850 &  0.2061 & -0.1001  \\  
   Variance & $1.9\gev$ &        &        &        &        & 1.0000 & 0.4473 &  0.2219 &  0.1436  \\
   Variance & $2.0\gev$ &        &        &        &        &        & 1.0000 &  0.5003 &  0.3342  \\
   Variance & $2.1\gev$ &        &        &        &        &        &        &  1.0000 &  0.6592  \\
   Variance & $2.2\gev$ &        &        &        &        &        &        &         &  1.0000  \\
   \hline
  \end{tabular}
   \caption{Correlation matrix (preliminary) for systematic errors on the 
   truncated first and
   second moments (four each, for different minimum \eg cuts).  This
   matrix is symmetric, so redundant entries are not shown. Column labels
   are shorthand for the identical row labels.  Systematic uncertainties
   included are those on the \BB subtraction, the energy scale (first
   moments only) and calorimeter resolution (second moments only).}
\label{table:moment_correlations_syst}
 \end{center}
\end{table*}

\begin{table*}[htbp]
 \begin{center}
  \begin{tabular}{|rr|cccccccc|}
   \hline
                                 &&  M  &  M  &  M  &  M  &  V  &  V  &  V  &  V  \\
   Quantity & Min. \eg  & 1.9 & 2.0 & 2.1 & 2.2 & 1.9 & 2.0 & 2.1 & 2.2 \\ \hline
   Mean     & $1.9\gev$ & 1.0000 & 0.9267 & 0.9486 & 0.8252 &-0.9057 &-0.9223  &-0.9234 &-0.7983  \\
   Mean     & $2.0\gev$ &        & 1.0000 & 0.9576 & 0.8669 &-0.7452 &-0.8183  &-0.8087 &-0.8960  \\   
   Mean     & $2.1\gev$ &        &        & 1.0000 & 0.9540 &-0.7390 &-0.7889  &-0.7843 &-0.7752  \\
   Mean     & $2.2\gev$ &        &        &        & 1.0000 &-0.5378 &-0.5857  &-0.5788 &-0.6023  \\  
   Variance & $1.9\gev$ &        &        &        &        & 1.0000 & 0.9650  & 0.9824 & 0.6983  \\
   Variance & $2.0\gev$ &        &        &        &        &        & 1.0000  & 0.9810 & 0.8035  \\
   Variance & $2.1\gev$ &        &        &        &        &        &         & 1.0000 & 0.8057  \\
   Variance & $2.2\gev$ &        &        &        &        &        &         &        & 1.0000  \\
   \hline
  \end{tabular}
   \caption{Correlation matrix (preliminary) for model-dependent errors on 
   the truncated first and
   second moments (four each, for different minimum \eg cuts).  This
   matrix is symmetric, so redundant entries are not shown. Column labels
   are shorthand for the identical row labels.  Uncertainties are
   included for the model-dependence of the signal efficiency and of the 
   correction for resolution, Lorentz boost and cutoff.  Sixteen
   models are averaged, as described in the text.}
\label{table:moment_correlations_model}
 \end{center}
\end{table*}

The correlation matrices allow a fit of the moments to any
parameterized theoretical calculation. However we note that of the
eight moments only five are independent. One way to see this is by
considering that there are two moments for the highest energy cut
($\egcms > 2.2 \gev$) and all the other moments can be derived by
adding the three lower energy bins. Thus in fitting one must choose no
more than five of the moments and check that the associated covariance
matrices are positive definite.

\subsection{\CP Asymmetry}
Within the Standard Model the 
combined direct \CP asymmetry for \bsg and \bdg decays,
\begin{equation}
  \acp = \frac{\Gamma(\b\to\s\g + \b\to\d\g)-
       \Gamma(\bbar\to\sbar\g + \bbar\to\dbar\g)}
       {\Gamma(\b\to\s\g + \b\to\d\g)+
       \Gamma(\bbar\to\sbar\g + \bbar\to\dbar\g)} \; ,
  \label{eq:CPdirect}
\end{equation}
is zero to a very good approximation \cite{ref:acppredict}. New physics
models with additional flavour violation can significantly enhance
\acp to a few percent \cite{ref:acptheory}. Furthermore, measurements
of the combined \acp in \bsg and \bdg processes complement those of
\acp in only \bsg \cite{ref:acp,ref:ksg,ref:acpbelle} to constrain new
physics models.

The use of lepton tagging on the non-signal \B for continuum
suppression allows a straightforward determination of \acp{}.
The lepton's positive (negative) charge tags the
signal side to contain a $b$ $(\bbar)$ quark. Therefore
\begin{equation}
 \acp = \frac{N^{+}-N^{-}}{N^{+}+N^{-}} \frac{1}{1-2\omega} \; ,
\end{equation} 
where $N^{+(-)}$ are the positive (negative) tagged signal yields and 
$1/(1-2\omega)$ is the dilution factor due to the mistag rate, $\omega$. 

The systematic uncertainty on the \BB background subtraction only
contributes a multiplicative uncertainty to the measured
\acp. Therefore, the statistical uncertainty dominates and we find the
optimal \egcms range for the \acp measurement to be 
$2.2<\egcms<2.7\gev$. The tagged signal yields are
$N^{+}=349\pm48$ and $N^{-}=409\pm 45$ where the uncertainties
are statistical. We use control samples of 
$\epem \to X \gamma$ and $\B \to X \piz,\eta$ to estimate 
the bias on \acp from any charge
asymmetry in the detector or in the \BB background. 
The bias is $-0.005\pm 0.013$, which is applied to the
measured asymmetry. The uncertainty on the measurement of the bias is taken
as an additive systematic uncertainty on the measured asymmetry. The
measured asymmetry, uncorrected for dilution, is $-0.084\pm 0.088 \pm 0.013$, 
where the first
uncertainty is statistical and the second systematic.

The mistag rate is $(11.9\pm 0.4)\%$. It is dominated by mixing in
\BzBzb events, which when averaged over \BzBzb and \BpBm events is
equal to half the time-averaged mixing rate $\chi_d/2 = (9.3\pm
0.2)\%$ \cite{ref:pdg}. The remaining $(2.6\pm0.3)\%$ is estimated from
simulation; these mistags are due to leptons from $D$ decay, $\pipm$
faking $\mu^{\pm}$, $\gamma$ conversions, \piz Dalitz decay, and charmonium
decay. The error is dominated by uncertainties in the modeling of
$D$ decay and fakes. The dilution factor of
$1/(1-2\omega)=1.31\pm0.01$ scales the measured \acp and its additive
uncertainties to give
\begin{equation}
  \acp  = -0.110 \pm 0.115 (\mathrm{stat.}) \pm 0.017 (\mathrm{syst.}) \;.
\end{equation}

The systematic uncertainty includes a 5.4\% multiplicative uncertainty  
from the \BB background subtraction and 1.0\% multiplicative uncertainty 
from the dilution factor. We have estimated that any model dependent 
uncertainty, due to differences in the \bxsg and \bxdg spectra, is 
much smaller than the systematic uncertainty.

\section{Conclusions}

\label{sec:conclusions}
We have performed a lepton-tagged fully-inclusive measurement of 
\bxsg decays.  We present a reconstructed photon
energy spectrum in the \Y4S frame, and partial branching
fractions above minimum reconstructed energies of 1.9, 2.0, 2.1 and
2.2 GeV (Table~\ref{table:partial_bfs_corrected}). We also
present the partial branching fractions, the
first and second moments of the true photon energy distribution in
the \B rest frame and the correlations between the first and second moments, 
above the same minimum photon energies
(tables~\ref{table:partial_bfs_corrected}, 
\ref{table:firstmom_datacorrected}, \ref{table:secondmom_datacorrected},
\ref{table:moment_correlations_stat}, \ref{table:moment_correlations_syst} 
and~\ref{table:moment_correlations_model}). Finally, we have used the charge 
of the tagging leptons to obtain the \CP asymmetry of the sum of 
\bxsg plus \bxdg with reconstructed photon energy above 2.2\gev.

We are grateful for the 
extraordinary contributions of our \pep2\ colleagues in
achieving the excellent luminosity and machine conditions
that have made this work possible.
The success of this project also relies critically on the 
expertise and dedication of the computing organizations that 
support \babar.
The collaborating institutions wish to thank 
SLAC for its support and the kind hospitality extended to them. 
This work is supported by the
US Department of Energy
and National Science Foundation, the
Natural Sciences and Engineering Research Council (Canada),
Institute of High Energy Physics (China), the
Commissariat \`a l'Energie Atomique and
Institut National de Physique Nucl\'eaire et de Physique des Particules
(France), the
Bundesministerium f\"ur Bildung und Forschung and
Deutsche Forschungsgemeinschaft
(Germany), the
Istituto Nazionale di Fisica Nucleare (Italy),
the Foundation for Fundamental Research on Matter (The Netherlands),
the Research Council of Norway, the
Ministry of Science and Technology of the Russian Federation, and the
Particle Physics and Astronomy Research Council (United Kingdom). 
Individuals have received support from 
CONACyT (Mexico),
the A. P. Sloan Foundation, 
the Research Corporation,
and the Alexander von Humboldt Foundation.


\begin{thebibliography}{99}
\def\mpl #1 #2 #3 {Mod.~Phys.~Lett.~{\bf#1},\ #2 (#3)}
\def\npb  #1 #2 #3 {Nucl.~Phys.~B~{\bf#1},\ #2 (#3)}
\def\plb  #1 #2 #3 {Phys.~Lett.~B~{\bf#1},\ #2 (#3)}
\def\pr   #1 #2 #3 {Phys.~Rep.~{\bf#1},\ #2 (#3)}
\def\prd  #1 #2 #3 {Phys.~Rev.~D~{\bf#1},\ #2 (#3)}
\def\prl  #1 #2 #3 {Phys.~Rev.~Lett.~{\bf#1},\ #2 (#3)}
\def\RMP  #1 #2 #3 {Rev.~Mod.~Phys.~{\bf#1},\ #2 (#3)}
\def\zpc  #1 #2 #3 {Z.~Phys.~C~{\bf#1},\ #2 (#3)}
\def\nim  #1 #2 #3 {Nucl.~Instrum.~Methods~{\bf#1},\ #2 (#3)}
\def\nima  #1 #2 #3 {Nucl.~Instrum.~Methods~A~{\bf#1},\ #2 (#3)}
\def\epjc #1 #2 #3 {Eur.~Phys.~J.~C~{\bf#1},\ #2 (#3)}
\def\rmp #1 #2 #3 {Rev.~Mod.~Phys~{\bf#1},\ #2 (#3)}
\def\npbps #1 #2 #3 {Nucl.~Phys.~B.~proc.~suppl~{\bf#1},\ #2 (#3)}
\def\progtp #1 #2 #3 {Prog.~Theo.~Phys~{\bf#1},\ #2 (#3)}
\def\etal{{\it et al.}}


\bibitem{ref:new-physics}
B.~Grinstein and M.B.~Wise, \plb 201 274 1988 ; 
W.S.~Hou and R.S.~Willey, \plb 202 591 1988 ;
J.L.~Hewett and J.D.~Wells, \prd 55  5549 1997 ;
T.~Hurth, \RMP 75 1159 2003 .

\bibitem{ref:br-theory}
P.~Gambino and M.~Misiak, \npb 611 338 2001 ;
A.J.~Buras, A.~Czarnecki, M.~Misiak and J.~Urban, \npb 631 219 2002 .

\bibitem{ref:theory-future} 
M.~Misiak and M.~Steinhauser, \npb 683 277 2004 . 





\bibitem{ref:kn}
A.L.~Kagan and M.~Neubert, \epjc 7 5 1999 .

\bibitem{ref:BBU}
D.~Benson, I.I.~Bigi and N.~Uraltsev, \npb 710 371 2005 .


\bibitem{ref:neubert}
M. ~Neubert, \epjc 40 165 2005 .

\bibitem{ref:N}
B. ~Lange, M.~Neubert and G.~Paz, hep-ph/0504071.

\bibitem{ref:cleobsg} CLEO Collaboration, S. Chen {\it et al.}, \prl 87
251807 2001 .



\bibitem{ref:bellebsg} BELLE Collaboration, P. Koppenburg {\it et al.},
\prl 93 061803 2004 .

\bibitem{ref:semi-inclusive}
\babar\ Collaboration, B.~Aubert {\it et al.}, contributed to the
$31^{\mathrm{st}}$ Intl. Conf. on High Energy Physics, ICHEP-02,
hep-ex/0207074.

\bibitem{ref:acptheory}
T.~Hurth, E.~Lunghi and W.~Porod, \npb 704 56 2005 . 

\bibitem{ref:AAH}
H.M.~Asatrian, H.H.~Asatryan and A.~Hovhannisyan, \plb 585 263 2004 .




\bibitem{ref:babar-nim}
\babar\ Collaboration, B.~Aubert {\it et al.}, \nima 479  1 2002 .




\bibitem{ref:geant}
GEANT4 Collaboration, D. Agostinelli {\it et al.}, \nima 506 250 2003 .

\bibitem{ref:jetset}
`PYTHIA 5.7 and JETSET 7.4: Physics and manual', by Torbj\"orn Sj\"ostrand (Lund U.), hep-ph/9508391.



\bibitem{ref:ksg} 
\babar\ Collaboration, B.~Aubert {\it et al.}, \prd 70 112006 2004 .


\bibitem{ref:semileptonic}
\babar\ Collaboration, B.~Aubert {\it et al.}, contributed to the 
$32^{\mathrm{nd}}$ Intl. Conf. on High Energy Physics, ICHEP-04,
hep-ex/0408075.

\bibitem{ref:DstFF}
\babar\ Collaboration, B.~Aubert {\it et al.}, contributed to the 
$32^{\mathrm{nd}}$ Intl. Conf. on High Energy Physics, ICHEP-04,
hep-ex/0409047.

\bibitem{ref:acppredict}
J.M.~Soares, \npb 367 575 1991 ,
T.~Hurth and T.~Manuel, \plb 511 196 2001 .


\bibitem{ref:acp}
\babar\ Collaboration, B.~Aubert {\it et al.}, \prl 93 021804 2004 .

\bibitem{ref:acpbelle}
BELLE collaboration, S.~Nishida {\it et al.}, \prl 93 031803 2004 .   

\bibitem{ref:pdg} 
Particle Data Group, S.~Eidelman {\it et al.}, \plb 592 1 2004 .


\end{thebibliography}
\end{document}